\documentstyle[aps]{revtex}
\begin{document}
\draft
\title{
Magnetotransport Scaling 
in Anomalous Metallic Phase 
near the Mott Transition
\cite{N}
}
\author{
O. Narikiyo
}
\address{
Department of Physics, 
Kyushu University, 
Fukuoka 810-8560, 
Japan
}
\date{
March, 2001
}
\maketitle
\begin{abstract}
The Hall angle in the doped V$_2$O$_3$ 
is discussed on the basis of an extended Ginzburg-Landau theory 
of spin fluctuations. 
A spinon-like collective transport process 
beyond the ordinary quasiparticle transport theory 
is necessary to understand the experiment. 
At the same time a criticism of a quasiparticle transport theory 
on the basis of the fluctuation exchange (FLEX) approximation is made. 
\end{abstract}
\vskip 18pt
  The magnetotransport scaling, the Hall angle, 
in the normal metallic state of the doped V$_2$O$_3$ 
indicates the existence of a spinon-like tranport process.
\cite{Rosenbaum} 
  Such a spinon-like transport was proposed 
in order to understand the magnetotransport scaling 
in the normal metallic state of 
the high-$T_{\rm c}$ cuprate superconductors.
\cite{Anderson} 
  Although the spinon is a well-defined elementary excitation 
only in one dimension, 
even in two or three dimensions 
the collective excitation of spin fluctuations 
has a spinon-like character. 
  In the itinerant-localized duality model 
a spinon-like transport process is naturally introduced 
in consistency with experiments 
\cite{spinon} 
in the metallic phase near the Mott transition 
where the Fermi surface of the renormalized quasiparticles 
has a nesting character. 
  This correlation induced nesting 
is inevitable when approaching to the Mott insulator 
and reduces the effective dimensionality. 
  In this Short Note 
we calculate the Hall angle 
on the basis of the itinerant-localized duality model 
for the metal near the Mott transition in three dimensions, 
which is relevant to the case of the doped V$_2$O$_3$, 
while the two-dimensional calculation 
relevant to the case of the high-$T_{\rm c}$ cuprate superconductors 
was performed some years ago. 
\cite{Hall_angle}

  The itinerant-localized duality model 
is an extension of the time-dependent Ginzburg-Landau theory 
for spin fluctuations. 
\cite{review} 

  The dynamical spin susceptibility $\chi({\bf Q+q},\omega)$is given by 
\begin{equation}
{\bar \chi}/\chi({\bf Q+q},\omega) 
= {\tilde r} + A q^2 - {\rm i}C\omega, 
\end{equation}
where ${\bf Q}$ is the characteristic wave vector 
of the antiferromagnetic spin fluctuations 
and ${\bar \chi}$ is the renormalized density of states 
at the Fermi level. 
  The quartic interaction of spin fluctuations is 
renormalized into the dimensionless distance ${\tilde r}$ 
from the critical point. 
  While $A$ and $C$ are constants in the ordinary 
Ginzburg-Landau theory, 
these should be dependent on the temperature and the doping rate
\cite{Hall_angle,3D} 
both of which are the measure from the phase boundary 
between the metal and the Mott insulator. 
  Approaching the phase boundary 
the spectral weight of the spin fluctuations 
is transferred from the low-energy itinerant part 
to the high-energy localized part 
in accordance with the development of antiferromagnetic correlations 
among local spins.
\cite{sim1,sim2} 
  This spectral weight transfer is one of the novel features 
in the metallic phase near the Mott transition 
not considered in the ordinary theory of the itinerant spin fluctuations. 
\cite{Moriya} 
  The reduction of the low-energy spectral weight 
leads to the so-called pseudogap phenomena. 
  At the same time the Fermi surface has a nesting character 
reflecting a fact that the spectral weight of the electrons 
also has large weights of local spin character at higher energies. 

  The resistivity $R$ is calculated using the above susceptibility as 
$R = R_0 + \Delta R$ where 
\begin{equation}
\Delta R \propto T^2 / A^{3/2} {\tilde r}^{1/2}. 
\end{equation}
  Here $R_0$ is the residual resistivity due to impurity scatterings 
and $\Delta R$ is the quasipartcle contribution. 
\cite{Kohno} 
  In the resistivity 
there is no collective contribution discussed below. 
  In order to explain experiments\cite{Rosenbaum} 
in the doped V$_2$O$_3$, 
$\Delta R$ should behave as $\Delta R \propto T^{3/2}$ 
so that $A^{3/2} {\tilde r}^{1/2} \propto T^{1/2}$ in the following. 

  The Hall coefficient is calculated as ref.\ 6 and 
the resulting Hall angle $\Theta_{\rm H}$ is given by 
$B \cdot \cot \Theta_{\rm H} = R / R_{\rm H}$. 
  The Hall coefficient $R_{\rm H}$ is given as 
$R_{\rm H} = R^\infty_{\rm H} + \Delta R_{\rm H}$ 
where $R^\infty_{\rm H}$ is 
the temperature-independent part of $R_{\rm H}$, 
the quasiparticle contribution, 
and $\Delta R_{\rm H}$ is the spinon-like contribution given by 
\begin{equation}
\Delta R_{\rm H} \propto 
{\hat R}^2 x^3 (x+\alpha)^2 / t^{1/2}
\equiv \Delta \rho, 
\end{equation}
where the dimensionless resistivity ${\hat R}$ 
is chosen as ${\hat R} = {\hat R}_0 + t^{3/2} $ 
and $x = 1/(t^2+h^2)$. 
  Here the temperature $T$ is normalized as $t \equiv T/\varepsilon_{\rm F}$ 
where $\varepsilon_{\rm F}$ is the renormalized Fermi energy, 
the energy scale for the deviation from the perfect nesting $H$ 
is normalized as $h \equiv H/2\pi\varepsilon_{\rm F}$ 
and the interaction between localized spins $J$ is normalized as 
$\alpha \equiv 2J/\pi a_f \varepsilon_{\rm F}$ 
where the dimensionless parameter $a_f$ is to be determined 
by the dispersion of the quasiparticle. 
  The experimental data
\cite{Rosenbaum} are well fitted
\cite{3D} 
if we assume that $\varepsilon_{\rm F} \sim $40K, $\alpha=10$ 
and $h=$0.1. 
  Hereafter we assume the same values. 

  Both $R^\infty_{\rm H}$ and $\Delta R_{\rm H}$ 
are contained in the linear response theory
\cite{linear_response}. 
  However, in the ordinary quasiparticle transport theory 
only the former is taken into account. 
  The latter, the collective contribution to the conductivity, 
appears in the presence of the magnetic field. 

  The Hall angle in the presence of the magnetic field $B$ 
is plotted as 
\begin{equation}
B \cdot \cot \Theta_{\rm H} \propto 
{\hat R} / ( \rho + \Delta \rho ), 
\end{equation}
where $R^\infty_{\rm H}$ is normalized as $\rho$. 
  The numerical results are shown in Fig.\ 2 
where the residual resistivity is chosen as ${\hat R}_0=3$ 
in accordance with experiments.
\cite{Rosenbaum} 
  These data are qualitatively compatible with the experiment 
where $B \cdot \cot \Theta_{\rm H}$ is roughly proportional to $T^2$. 
  Such a temperature dependence is unexpected 
from the ordinary quasiparticle transport theory. 
  Since our estimation is too crude, 
we need further elaboration to achieve a quantitative agreement. 

  In conclusion, 
we have demonstrated the existence of the spinon-like 
transport process in the doped V$_2$O$_3$. 
  It is a straightforward application of 
the itinerant-localized duality model, 
which is effective to understand observed anomalies 
in the two-dimensional metallic state of 
the high-$T_{\rm c}$ cuprate superconductors, 
to the three dimensional case. 

  As pointed out in the previous studies
\cite{Rosenbaum,Anderson}, 
the spinon-like transport process is necessary 
to explain the Hall angle data. 
  On the contrary 
there is an attempt to derive an anomalous temperature 
dependence of the Hall conductivity 
in the normal metallic state of 
the high-$T_{\rm c}$ cuprate superconductors 
from the vertex correction 
in the fluctuation exchange (FLEX) approximation.
\cite{KKU,cond-mat} 
  Such an attempt should be criticized, 
since the theoretical result is not compatible 
with experiments by the reason in the following. 
  The main issue here is the relation between 
the temperature dependences of the Hall conductivity 
and the spin-lattice relaxation rate $1/T_1$ in underdoped region. 
  The case of overdoped region is trivial, 
since the temperature dependence of the Hall conductivity 
is weak and easily ascribed to the anisotropy of the life time 
of the quasiparticle on the Fermi surface. 
  In underdoped region 
$1/T_1 T$ measured by the NMR shows the so-called psuedogap behavior, 
while the Hall conductivity is enhanced. 
  In YBCO-60K case
\cite{Y1,Y2} 
such behavior is observed for 100K $< T <$ 150K. 
  Sudden decrease of $R_{\rm H}$ for $T <$ 100K 
is ascribed to the superconducting fluctuation. 
  In HgBaCaCu1212-125K case
\cite{Hg1,Hg2} 
such behavior is observed for 140K $< T <$ 200K. 
  Sudden decrease of $R_{\rm H}$ for $T <$ 140K 
is ascribed to the superconducting fluctuation. 
  These facts contradict with the FLEX approximation, 
since the vertex correction in the theory 
dominated by the Maki-Thompson process proprotional to 
the ${\bf q}$-integrated weight of the spin fluctuation 
Im$\chi({\bf Q+q},\omega)/\omega$ 
in the limit of $\omega \rightarrow 0$, 
which is nothing but the measured quantity in the NMR experiment 
and has to show the pseudogap behavior, 
shows a monotonous temperature dependence. 
  Both the Hall conductivity and the spin-lattice relaxation rate 
should be understood in a single framework of a spin-fluctuation theory 
in the temperature range of the present issue 
and it is possible.
\cite{review,sim1,sim2} 
  The inconsistency of the FLEX approximation 
results from the violation of the Pauli exclusion principle 
in the approximation. 
  Thus in this approximation 
the spectral weight transfer discussed above 
and crucial to understand the pseudogap phenomena 
is not correctly described. 
  This crack of the FLEX approximation 
has been demonstrated in the context of the formation of 
the high-energy Hubbard band for local interaction models. 
\cite{Hubbard_band}
  At the same time we can not obtain the correct 
transport equation, the Boltzmann equation, 
if the Pauli principle is violated, 
since the collision term is highly restricted by the principle. 

\newpage

\vskip 100pt

\begin{figure}
\caption{
Feynman diagrams for the Hall coductivity. 
The quasiparticle transport process is shown in (a) 
and the collective one in (b). 
The triangle vertexes represent the appropriate vertex corrections. 
}
\label{fig:1}
\end{figure}

\begin{figure}
\caption{
Temperature dependence of the Hall angle.
}
\label{fig:2}
\end{figure}

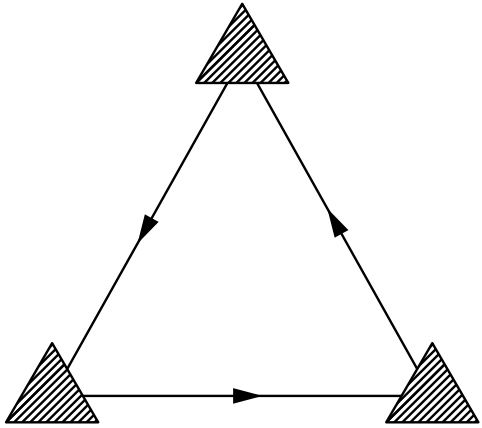
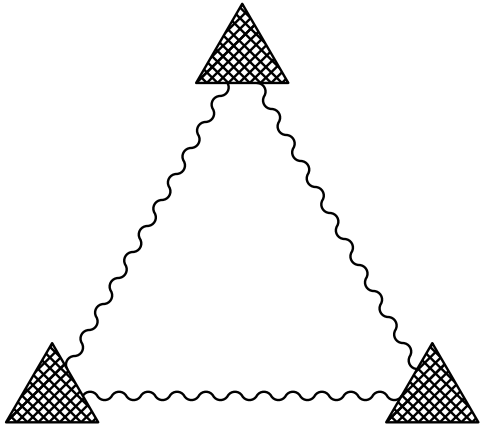
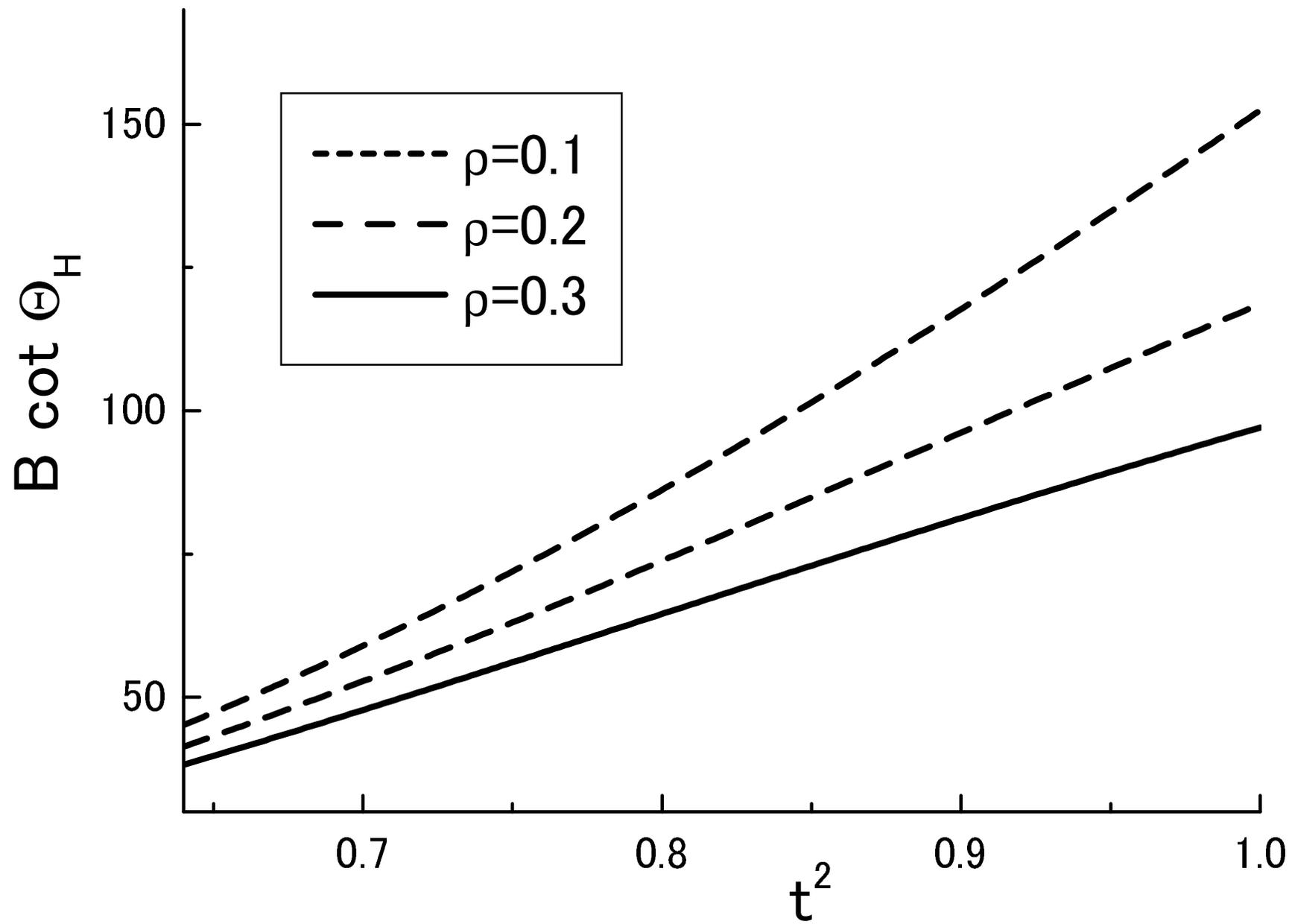
\end{document}